# Optical transparency and electrical conductivity of single-wall carbon nanotubes and of intermediate filaments of porcine Müller cells

Igor Khmelinskii,[1] and Vladimir Makarov[2]

[1]Universidade do Algarve, FCT, DQB and CIQA, 8005-139, Faro, Portugal

[2]University of Puerto Rico, Rio Piedras Campus, PO Box 23343, San Juan, PR 00931-3343, USA

*Corresponding Author*: Dr. Vladimir I. Makarov

University of Puerto Rico, Rio Piedras Campus, PO Box 23343, San Juan, PR 00931-3343, USA

*Phone*: (1)787-529-2010

E-mail: vmvimakarov@gmail.com

## *Abstract*

In the present study, we continue investigation of the high-contrast vision in the inverted retina of the vertebrates eyes. We report a method of separation and purification of porcine (*Sus scrofa domestica*) intermediate filaments (IFs), extracted from the retinal Müller cells (MCs). We also report experimental and theoretical methods of measurements and calculations of the reduced resistivity and light transmission by the IFs and single-wall carbon nanotubes (SWCNTs). The measured reduced resistivity values were $(3.1\pm0.3)\times10^{-4}$ and $(2.8\pm0.2)\times10^{-4}$ $\Omega\cdot m^{-1}\cdot cm^{2}$, respectively, being quite close to those of typical metals. We report a method for measuring the light energy transmission by the intermediate filaments and single-wall carbon nanotubes. We found that these structures efficiently transfer light energy along its axis, with the light reemitted at the other end of the structure. We also report spectral selectivity of the IFs. The reported results demonstrate that the assumptions we made in deducing the theory of high-contrast vision in an inverted retina were correct and fully supported by the presently reported experimental results.

## I. Introduction

The light transmission by the inverted retina of the vertebrates has already been discussed for more than 100 years. The idea of light transmission by glial Müller Cells (MCs) of the inverted retina has been proposed more than 10 years ago [1,2], being now considered the main mechanism of light transmission by the inverted retina to the photoreceptor cells. It was shown by Zueva, et al. (1916) [3] that the Intermediate Filaments (IFs) traverse the entire length of MCs; and it was later proposed that IFs are in fact the structures that conduct the light energy [4]. Note that these IFs may be sensitive to light polarization and have a limited spectral transmission [4,6]. We have also assumed that IFs located in the MCs should be electrically conductive, which allowed to conveniently account for their capacity to transfer the light energy in the retina. Unfortunately, no experimental data on the light energy transmission by the IFs were reported till presently, the same applies to the information on their electric conductivity.

Structure and properties of the IFs in different biological systems were extensively studied earlier [7-12]. In 1959, an unusual filamentous polymer, now called a beaded filament, was described in the lens of the eye. The association of lenticular chaperones, α-crystallins, with the filament contributes to the characteristic beaded morphology, giving also important clues to the function of this unusual filament in the lens [7,8]. In the review paper [9], the two separate intermediate filament systems present in the eye lens were discussed. Canonical 8-11 nm IFs composed of Vimentin are assembled in the lens epithelial cells and younger fiber cells, while the fiber cell – specific beaded filaments start accumulating as the fiber cell elongation initiates. The basic methods of purification of the important lens fractions and the analysis of the distribution of IF proteins in the eye lens were reported by Perng, et al. (2004) [10]. We used their methods [10] with some modifications in the current study. The functions of the IFs were extensively discussed in the reports by Strelcov, et al. (2002, 2017) [11,12]. They demonstrated [11,12] that IFs not only have the cytoskeletal function, but many other physiological functions. However, electric conductivity and light energy transmission functions of the MC IFs have not been studied yet.

As we already noted, quantum mechanism of the light energy transport by the MC IFs was proposed earlier [4]. In the frameworks of this mechanism, IFs absorb photons, passing to an electronically excited state, represented by a wave package (exciton) of quasiresonance excited states distributed over the entire length of the IF. Such wave package propagates along the IF, and upon arrival to the other end of the IF may again be transformed into a photon. This allows the IFs to transmit light energy with an efficiency close to unity, within a certain wavelength range for the given IF geometric parameters [4-6]. The quantum mechanism [4-6] requires electrically conductive IFs, which greatly simplifies the quantum mechanical treatment of such macromolecular systems as IFs, built of numerous protein molecules.

Regarding the proposed quantum theory [4] for the light energy transmission by the IFs, the systems that are closest to the IFs in their properties are carbon nanotubes (CNTs). In particular, properties of single-wall CNTs (SWCNTs) should admit the most straightforward interpretation. Therefore, here we turned to SWCNTs and their electrical and optical properties, as a model for the more complex IFs. Note that $\pi$-conjugated carbon systems are electric superconductors, typical examples being SWCNTs and graphene [13]. Since SWCNTs are axisymmetric with electrically conductive walls, they made a reasonable model for our IF waveguides. Note that SWCNT is a hollow tube made of a one-atomic-layer-thick graphene sheet. The wrapping direction may be described by the chirality vector (***n, m***), its coordinates denoting the number of unit vectors along the two directions of the graphene sheet crystal lattice [14]. Thus, a tube with $n = m$ (chiral angle = 30°) is an armchair tube, while a tube with $m, n \gg 1$ and $|m - n| = 3k$ has metallic conductivity [15]. The diameter $d$ of the SWCNT is given by

$$d = \frac{c}{\pi}\sqrt{(n+m)^2 - mn}, \qquad (1)$$

where $c$ = 0.246 nm [13]. The optical absorption spectrum of SWCNTs includes several electronic transitions: $v_2 \to c_2$ ($E_{22}$) or $v_1 \to c_1$ ($E_{11}$), etc. [15-17]. These transitions are relatively sharp and may identify nanotube types. Interestingly, the plasmon line in the absorption spectrum of the SWCNT with $d$ = 12 nm is at ca. 4.5 eV [15]. Although it is obvious that the diameter is the most important property of the structure, we cannot expect the properties of the IFs to exactly coincide with those of SWCNTs, even for the same diameter. However, taking into account the axial symmetry of both the IFs and SWCNTs, along with the supposed electric conductivity of their walls, they both may be expected to efficiently transmit light by the quantum mechanism. We have studied these phenomena in detail earlier [4-6].

As the presently studied properties of the IFs and SWCNTs [3-6] have never been reported, here we carried out systematic measurements of their electric conductivity and light energy transmission efficiency. These measurements used specially designed and built electrode and capillary matrices. This allowed us to measure the resistance of a single IF and SWCNT, as well as the light energy transmission along their axis. We report the average reduced resistance of the IFs and SWCNTs of $(3.1\pm0.3)\times10^{-4}$ and $(2.8\pm0.2)\times10^{-4}$ $\Omega\cdot m^{-1}\cdot cm^2$, respectively, and also the values for their light transmission efficiency. Note that the reduced resistivity of the studied macromolecular systems is comparable with that of metals [18].

## II. Materials and methods

In this section, the materials and methods used in the current study will be described in detail.

a) Isolating IFs from porcine eyes and their characterization

The following procedure was used for isolating the IFs. Fresh porcine eyes were obtained from Toa Bajo Meat Factory (Puerto Rico). These were cooled to +4°C and transported to the laboratory in a commercial refrigerator, at the same temperature. Four retinas were removed from the eyes, totaling 0.523 mg. This material was frozen to -70°C during 5 min and finely chopped mechanically. Chopped homogeneous material was warmed to 20°C and mixed with 10 ml bidistilled water. 10 mg of the microcrystalline diamond (Sigma-Aldrich) was added to the mixture and ultrasonicated for 30 min at +20°C. The mixture was filtered using paper filter

(Sigma Aldrich) with pore diameter of 500 nm, the liquid fraction was filtered again using a paper filter with 20 nm pore diameter. The second filter was placed into a glass beaker with 10 ml of bidistilled water, and mildly ultrasonicated for 5 min at $+20^{\circ}C$. The suspension was centrifuged for 15 min at 25000 g on a Sorvall RC-5B, the fraction of IFs with the mass $150\ k \leq M \leq 300\ k$ was removed and used in further experiments. The selected fraction was diluted to the total volume of 10 ml and stored at $+4^{\circ}C$. The estimated total yield of the IFs was about 32 mg.

The samples were characterized using high-resolution transmission electron microscopy (HRTEM) on JEM-1400 Plus apparatus, spectrophotometry (Shimadzu UV-3600 Plus Spectrophotometer), spectrofluorimetry (F-7100 Hitachi Spectrofluorimeter). The method for the TEM sample preparation was described earlier [3]: 1 ml of the diluted fraction was mixed with 1 ml of the fixative: 2.5% glutaraldehyde, 4% paraformaldehyde in 90 mM sodium cacodylate buffer with 0.02 mM $CaCl_2$, pH 7.2-7.4, and stored for a month at +4°C. The suspension was filtrated, the filter was ultrasonicated with 1 ml of aqueous solution containing 90 mM sodium cacodylate buffer, and suspension was postfixed with 1 ml of 1% osmium tetroxide ($OsO_4$) with 1.5% $K_4[Fe(CN)_6]$ for 30 minutes in the same buffer, filtrated, the filter was washed, ultrasonicated in 1 ml of 1% $OsO_4$ solution and incubated in this solution for 30 min, filtered the filter was washed in bidistilled water, ultrasonicated in 1 ml of 2% aqueous uranyl acetate $[UO_2(CH_3COO)_2 \cdot 2H_2O]$ and the suspension was incubated for 1h. Next the suspension was filtered, the filter washed with bidistilled water and ultrasonicated in 1 ml of bidistilled water. One drop (ca. 0.03 ml) of the homogeneous suspension was transferred to TEM carbon grid, and then fixed by EMbed 812 EMS epoxy resin (all chemicals were obtained from Electron Microscopy Sciences, Hatfield, PA, USA).

The TEM image of the sample is shown in Figure 1a.

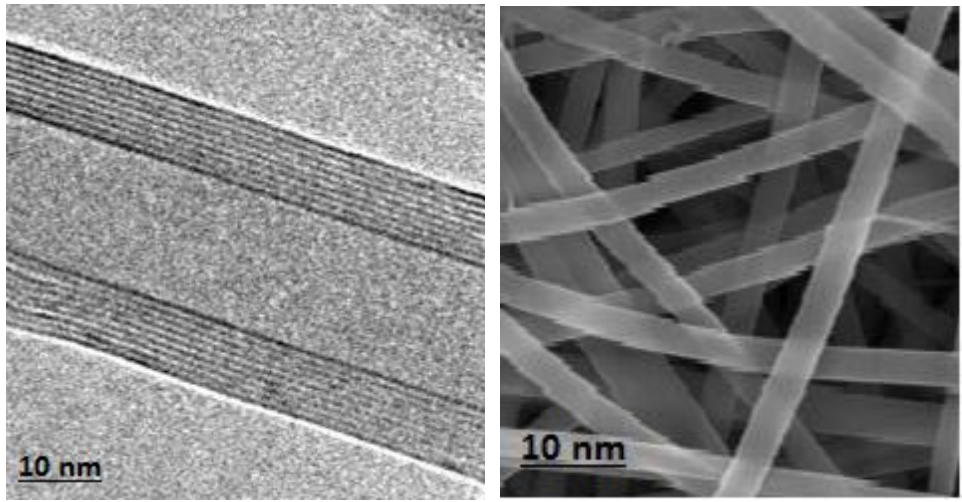

Figure 1. The TEM image of the IF sample prepared by procedure described above, where porcine retina was used for the sample preparation (a). The TEM of the SWCNTs prepared by procedure described below

Figure 1a shows that the IF diameter is close to 10.2 nm, in agreement with earlier reports [3]. The estimated average IF length is 117.3 μm. Thus, taking into account that IFs of the porcine retina were properly isolated, their suspension was further used for measurements of electrical and light conductivity of the intermediate filaments.

b) SWCNT samples

Commercial single-wall carbon nanotubes (SWCNT) from Firstnano Inc. (www.firstnano.com) were used without additional purification. The average length of the SWCNTs was 7.8 μm, and their average diameter 4.5 nm. To produce hydrophilic SWCNTs, 100 mg of commercial SWCNTs were boiled in 30 ml of 30% $HNO_3$ during 30 min. The sample was filtered and washed three times by 50 ml of bidistilled water. The filter was ultrasonicated for 15 mins in 15 ml of bidistilled water, and the SWCNT suspension was stored at room temperature.

One drop (ca. 0.03 ml) of the suspension was put on a carbon TEM grid, dried, and heated during 30 min at 120°C. The TEM image of SWCNTs is shown in Fig. 1b, and their suspension was used in further experiments.

c) Electric conductivity measurements

The matrix of the gold electrodes consists of 500 pairs of electrodes. The matrix was produced using electron beam lithography (JBX-6300FS, JEOL), with its electrodes connected between each other as shown in Fig. 2a.

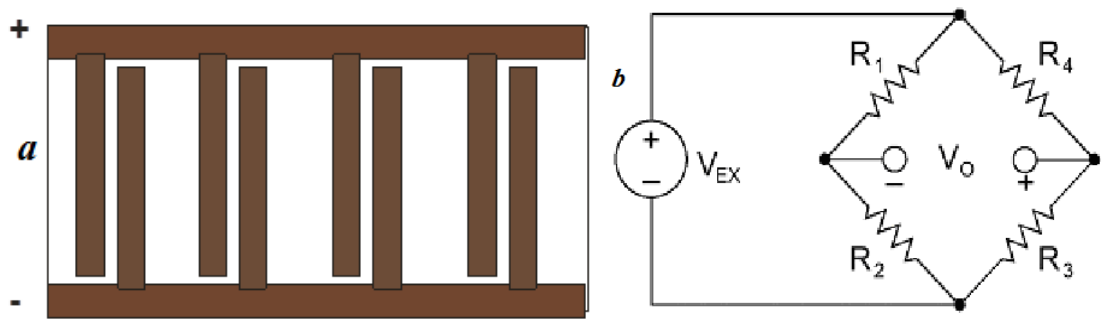

Figure 2. Schematic representation of the matrix of the golden electrodes (*a*) and measurement scheme (*b*).

Each electrode was 1 μm wide, the distance between the electrodes in a pair was 20 nm and the distance between two neighboring electrode pairs was 20 μm. The total size of the matrix active area was 10×20 mm$^2$. Total electrode sueface is about 80% from rotal EM surface.

The electrode matrix was connected into the Wheatstone bridge (WB; ESI Wheatstone system, 230b&DC Detector 803a), as shown in Figure 2b. In the WB, $R_1 = R_2 = R_3 = 1$ MΩ, $R_4$ – variable resistor to balance the bridge; the electrode matrix was connected in parallel to $R_3$. The signal from WB was amplified by an amplifier (KEITHLEY 6430 Source Measure Unit) connected to a PC computer via GPIB board (National Instruments Inc., QUANCOM PCI GPIB Card). The data acquisition system was controlled by the software running LabView environment. The measurement circuit permitted to measure currents down to 0.1nA. All experiments were carried out at 1.000 V DC applied to the Wheatstone bridge, provided by the KEITHLEY 6430 unit.

The electrode matrix was mounted inside a Pyrex sample cell with 4.5 mL total volume that contained either bidistilled water or the sample suspension, and the cell was mounted into a water thermostat with temperature controllable in the 5-40 °C interval. The bridge was balanced with bidistilled water in the sample cell (note that the electrode matrix resistance in pure water was much larger than 1 MΩ).

d) Light transmission by single SWCNT and single IF

This set of experiments, with their prototype described earlier [19-21], was also carried out here. Earlier a nanocapillary matrix was described [19], which was used to decode the nucleotide sequence in polynucleotides. Our capillary matrices have a different structure compared to those described earlier [18-22]. They were produced as follows. A silicon substrate (Wafer Word Inc,) sized 12.5×12.5×0.15 $mm^3$ was ultrasonicated for 20 min in 20 ml of isopropanol with 20 mg of commercial diamond powder (Sigma-Aldrich). The substrate was next washed with acetone, isopropanol and deionized water, and dried at 100ºC. Electron beam-lithography system was next used to produce $10^4$ through holes in a 100×100 matrix, each 15 nm in diameter (M1 matrix, used with IFs) or 5 nm diameter (M2 matrix, used with SWCNTs). The holes were spaced at 20 µm intervals, covering a 2×2 $mm^2$ square. The capillary matrices were mounted into round aluminum frames 25 mm in diameter, which were in turn mounted into Pyrex sample cells with fused silica windows as shown in Figure 3a.

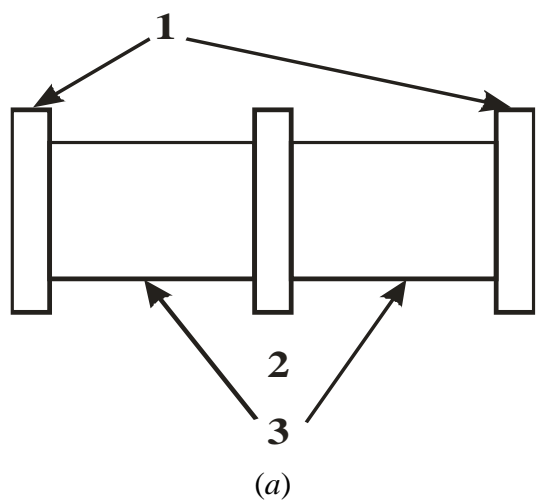

(*a*)

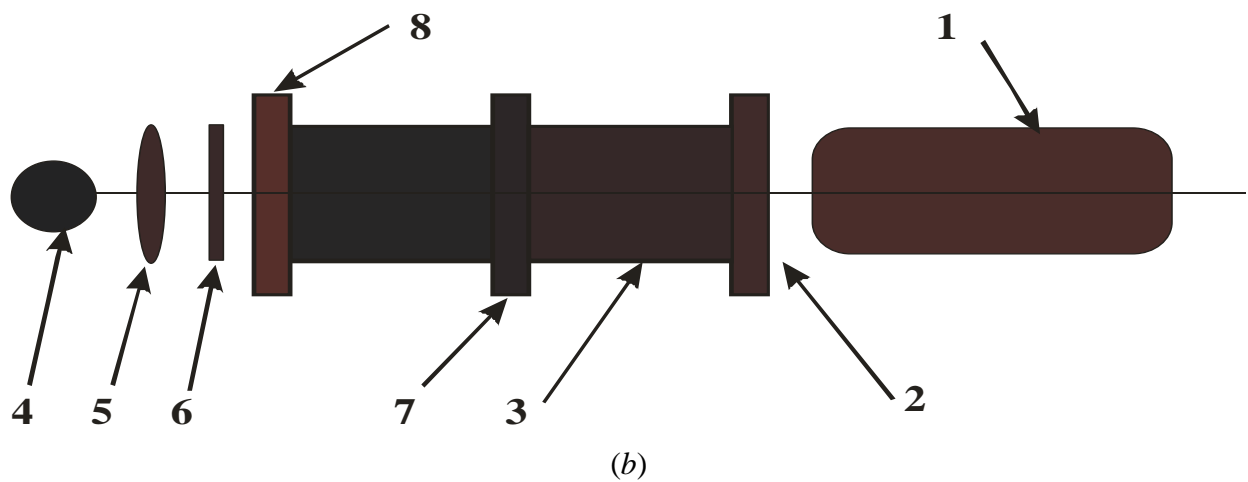

(b)

Figure 3. Schematic presentation of the cells of the capillary matrixes: (1) quartz windows; (2) capillary matrix frame with capillary matrix; (3) pyrex tubes with diameter 20 mm and length of 25 mm (pyrex tubes with internal diameter 8 mm, which are connected to the body both tubes of the cell are not shown in Figures 3a and 3b) (a) and system used for measurements of the light transmission measurements by the IFs and SWCNTs: all elements of this system are clarified below (b).

As seen in Figure 3a, a capillary matrix frame separate the cell into two equal sections, each 29 mm in diameter and 5 mm thick. The assembly used to measure the light transmission by the IFs and SWCNTs is shown in Figure 3b. It includes a photomultiplier (Hamamatsu, PMT Module No. 57-561, item 1) that was looking at the output window (2) of the cell (3). A high-pressure 100W mercury lamp (4) was used as the light source (ESI 1200 100W MSR Lamp; Planet Bulb Inc.), its radiation was focused by appropriate lenses (5), filtered by a water-cooled glass filter (6) to isolate the 404.7, 435.8 and 546.1 nm Hg emission bands, and focused onto the 2×2 $mm^2$ square of the capillary matrix (7). Some of the light was transmitted by the capillary matrix to the other window (8) of the cell. The two sections of the cell were isolated optically; therefore, light could only get to the PMT if it passed through the matrix. A photon counter (SR400, Stanford Research) data acquisition system was used in these optical experiments.

We also recorded time-resolved emission using frequency-doubled radiation of dye lasers for excitation (LPD-2000, Λ-Physics with BBO crystal, Λ-Physics). Coumarin-4 dye was used in the 370-580 nm range (frequency-doubled for 220-290 nm radiation), pumped by the fourth harmonics (at 266 nm) of a YAG laser (Surelight-II, Continuum Inc.). The laser pulse duration was 7-10 ns. The dye laser radiation was defocused onto the sample cell volume. The emission was collected by a 30-cm spherical $CaF_2$ lens and detected by a photodiode (PD1; DET10A Biased Si Detector from THORLABS) or a photomultiplier (PMT-H9305-03, Hamamatsu), after passing through a neutral density filter. The data acquisition system contained a PC computer, a digital oscilloscope (WaveSurfer 400 series, LeCroy), two digital delay generators (DG-535, Stanford Research), a photo-detector (PD: DET10A Si Detector from THORLABS), two boxcar integrators (SR-250, Stanford Research), a fast amplifier (SR-240, Stanford Research), and a computer interface board (SR-245, Stanford Research). For the spectroscopy experiments, the emission signal was recorded by the digital oscilloscope and averaged for a certain number of laser pulses (typically 5 laser pulses per frequency step). Control of the output radiation energy of the probe laser and monochromator scan operation was carried out by the PD and PCI-6034E DAQ I/O board (National Instruments), with the control code in the LABVIEW environment running on a second Dell PC. The presently used experimental methods allowed recording the time evolution of the emission with 2.5 ns resolution.

## III. Results and discussion

*a) Absorption and emission spectra of samples*

Absorption and emission spectra for both sample suspensions were measured using a Hitachi U-3900H UV-Visible Spectrophotometer and Edinburgh Instruments FS5 Spectrofluorometer. The absorption spectra are shown in Fig. 4.

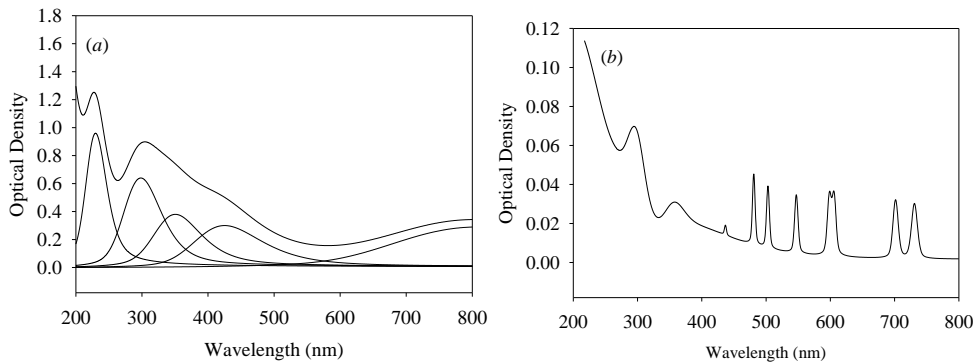

Figure 4. Absorption spectra of the IF suspension (*a*) and SWCNT suspension (*b*). Deconvoluted transition bands in absorption spectrum of IFs are as well presented in Fig. 4a.

The absorption spectrum of SWCNTs was discussed in detail earlier [23], including the assignment of the spectral bands [23], and therefore will not be discussed here. The absorption spectrum of the aqueous suspension of the IFs was deconvoluted, with at least five separate bands present, as also shown in Fig.4a. The positions and widths of the respective bands are listed in Table 1.

Table 1. Maxima and widths of the individual deconvoluted bands in the absorption spectrum of the IFs (Fig. 4a)

| Band Number | Band Maximum, nm | Band Maximum, cm$^{-1}$ | Bandwidth, cm$^{-1}$ |
|---|---|---|---|
| 1 | 809 | 12352 | 3068 |
| 2 | 425 | 23522 | 3875 |
| 3 | 350 | 28521 | 3971 |
| 4 | 298 | 33526 | 4374 |
| 5 | 229 | 43524 | 4077 |

Electric conductivity and absorption bands in UV and visible range lead us to an assumption that the properties of the IFs, similar to the nanotubes, depend on the conjugated $\pi$-system existing either in separate protein molecules composing the IFs, or comprising entire IFs.

The emission spectra for both IFs and SWCNTs at 220 nm excitation are shown in Fig. 5.

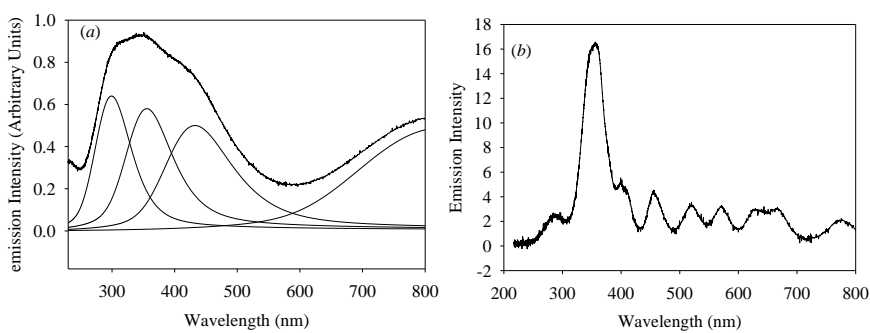

Figure 5. Emission spectra of IFs (*a*) and SWCNTs (*b*) excited by radiation with 220 nm wavelength. Deconvoluted transition bands in emission spectrum of IFs are as well presented in Fig. 5a.

The emission spectrum of SWCNTs was studied and analyzed earlier [23], and will not be discussed here. Note that strong emission transition band in SWCNTs is observable at 355 nm wavelength. Apparently, this transition band can be assigned to a $\pi^* \to \pi$ electronic transition. As regards the IF emission spectrum, spectral deconvolution produced several emission bands, with their parameters listed in Table 2.

Table 2. Positions of the maxima and widths of the individual bands deconvoluted from the IF emission spectrum (Fig. 5a)

| Band Number | Band Maximum, nm | Band Maximum, cm$^{-1}$ | Bandwidth, cm$^{-1}$ |
|---|---|---|---|
| 1 | 829 | 12052 | 3071 |
| 2 | 432 | 23122 | 3877 |
| 3 | 355 | 28121 | 3975 |
| 4 | 299 | 33423 | 4367 |

The positions of the emission band maxima and the bandwidths are quite close to the respective parameter values obtained of the absorption spectrum. There is a small red shift of the emission band maxima in comparison with the respective band maxima in the absorption spectrum. Therefore, the respective electronic transitions are vertical transitions with good accuracy, with any relaxation playing only a minor role.

Next we present the results on the IF and SWCNT emission dynamics. Both samples were excited by pulsed laser radiation at 225 nm, with the total emission detected at wavelengths exceeding 300 nm. The emission decay traces are shown in Fig. 6.

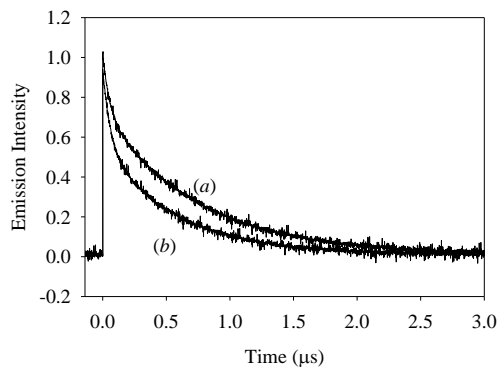

Figure 6. Emission decays of IF sample (*a*) and SWCNTs (*b*) excited by pulsed radiation with 225 nm wavelength.

The recorded decay traces were fitted by biexponential functions:

$$I_{Emis}(t) = A_1 e^{-\frac{t}{\tau_1}} + A_2 e^{-\frac{t}{\tau_2}} \qquad (2)$$

where $A_i$ ($i = 1, 2$) is the amplitudes of *i-th* exponential function and $\tau_i$ is the characteristic decay time of the *i-th* exponential function. The values of the respective decay times are listed in Tab. 3.

Table 3. Characteristic times of the emission signals of IF and SWCNT samples.

| Sample | Decay time $\tau_1$, ns | Decay time $\tau_2$, ns |
|---|---|---|
| IF | 63.9±1.8 | 795.1±15.7 |
| SWCNT | 74.5±1.3 | 650.1±14.2 |

The measured characteristic times will be compared to the respective times of light energy transmission by IFs and SWCNTs.

*b) Measurement of the IF and SWCNT resistance*

The set of experiments of IF and SWCNT resistance measurements was carried out at different number densities of IFs and SWCNTs, while the temperature varied in the 5 – 40°C range. The number density of the original sample suspensions of both IFs and SWCNTs was estimated taking into account the TEM images and the drop volume, which was 0.03 ml (see above). Thus, the estimated number density of the IFs in the suspension was $(2.1\pm0.3)\times10^8$ cm$^{-3}$, while that of the SWCNTs was $(3.6\pm0.5)\times10^9$ cm$^{-3}$.

Measurements were carried out at different temperatures and different concentrations of IFs and SWCNTs. The measurement temperatures were 5, 15, 25, 35 and 40°C, while the concentration of the IFs varied in the range of $(2.1\pm0.3)\times10^7$ to $(2.1\pm0.3)\times10^8$ cm$^{-3}$ with $(2.1\pm0.3)\times10^7$ cm$^{-3}$ step, and that of SWCNTs, in the $(3.6\pm0.5)\times10^8$ to $(3.6\pm0.5)\times10^9$ cm$^{-3}$ range with $(3.6\pm0.5)\times10^8$ step. The output signal of the Wheatstone bridge was recorded starting from the moment the sample suspension was added to the sample cell. Typical kinetic traces of these transient signals are shown in Fig. 7.

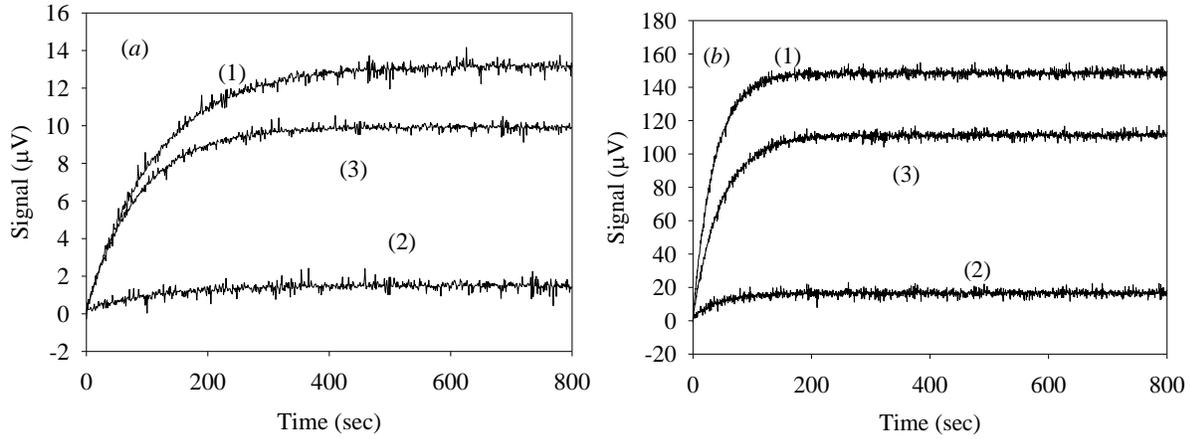

Figure 7. Time dependences of signal measured for (a) IFs (1) the $(2.1\pm0.3)\times10^8$ cm$^3$ at 5°C, (2) $(2.1\pm0.3)\times10^7$ cm$^{-3}$ at 5°C and (3) $(2.1\pm0.3)\times10^8$ cm$^3$ at 40°C, and (b) SWCNT (1) $(3.6\pm0.5)\times10^9$ cm$^{-1}$ at 5°C, (2) $(3.6\pm0.5)\times10^8$ cm$^{-3}$ at 5°C and (3) $(3.6\pm0.5)\times10^9$ cm$^{-3}$ at 40°C.

The recorded transient kinetics were fitted by an exponential function, with the fitting parameters $A$ and $k$ listed in Table 4.

$$V(t) = A(1 - e^{-kt}) \qquad (3)$$

Table 4. Values of the $A$ and $k$ fitting parameters to the transient output signal of the resistance measurement circuit, after adding the sample suspension to the Pyrex cell.

**IFs**

| $n_{IF}\times10^{-7}$, cm$^{-3}$ | 2.1 | 4.2 | 6.3 | 8.4 | 10.5 | 12.6 | 14.7 | 16.8 | 18.9 | 21.0 |
|---|---|---|---|---|---|---|---|---|---|---|
| $T = 278$ K | | | | | | | | | | |
| $A$, µV | 1.50 | 2.95 | 4.10 | 5.98 | 6.93 | 8.27 | 9.51 | 10.77 | 12.28 | 13.20 |
| $k\times10^3$, s$^{-1}$ | 9.0 | 8.6 | 9.2 | 8.9 | 8.8 | 8.9 | 9.1 | 8.8 | 8.7 | 8.9 |
| $T = 288$ K | | | | | | | | | | |
| $A$, µV | 1.37 | 2.61 | 3.79 | 5.08 | 6.19 | 7.18 | 8.40 | 9.74 | 10.81 | 12.06 |
| $k\times10^3$, s$^{-1}$ | 9.7 | 9.8 | 9.6 | 9.4 | 9.9 | 9.7 | 9.8 | 9.7 | 9.9 | 9.8 |
| $T = 298$ K | | | | | | | | | | |
| $A$, µV | 1.26 | 2.42 | 3.55 | 4.49 | 5.65 | 6.75 | 7.85 | 8.85 | 10.02 | 11.09 |
| $k\times10^3$, s$^{-1}$ | 10.6 | 10.1 | 10.8 | 10.6 | 10.6 | 10.3 | 10.9 | 10.5 | 10.4 | 10.9 |
| $T = 308$ K | | | | | | | | | | |
| $A$, µV | 1.17 | 2.23 | 3.33 | 4.25 | 5.16 | 6.26 | 7.27 | 8.17 | 9.20 | 10.25 |
| $k\times10^3$, s$^{-1}$ | 11.5 | 11.4 | 11.8 | 11.3 | 11.7 | 11.6 | 11.4 | 11.3 | 11.6 | 11.5 |
| $T = 313$ K | | | | | | | | | | |
| $A$, µV | 1.12 | 2.13 | 3.11 | 3.94 | 5.04 | 5.99 | 6.90 | 7.86 | 8.92 | 9.87 |

| $k\times10^3$, s$^{-1}$ | 11.9 | 12.0 | 12.1 | 11.8 | 11.8 | 11.9 | 11.7 | 12.2 | 11.9 | 11.8 |

**SWCNTs**

| $n_{CNT}\times10^{-8}$, cm$^{-3}$ | 3.6 | 7.2 | 10.8 | 14.4 | 18.0 | 21.6 | 25.2 | 28.8 | 32.4 | 36.0 |
|---|---|---|---|---|---|---|---|---|---|---|
| $T = 278$ K | | | | | | | | | | |
| $A$, μV | 16.51 | 28.93 | 47.36 | 60.95 | 75.56 | 91.12 | 105.87 | 115.87 | 134.71 | 148.44 |
| $k\times10^2$, s$^{-1}$ | 2.81 | 2.83 | 2.79 | 2.76 | 2.87 | 2.81 | 2.85 | 2.78 | 2.85 | 2.82 |
| $T = 288$ K | | | | | | | | | | |
| $A$, μV | 15.38 | 30.73 | 43.50 | 58.09 | 73.43 | 85.23 | 98.74 | 113.21 | 126.77 | 138.32 |
| $k\times10^2$, s$^{-1}$ | 3.02 | 3.08 | 2.98 | 2.93 | 3.06 | 3.01 | 3.05 | 2.98 | 3.03 | 3.05 |
| $T = 298$ K | | | | | | | | | | |
| $A$, μV | 14.43 | 29.42 | 40.06 | 54.69 | 68.83 | 81.81 | 91.20 | 105.66 | 119.91 | 129.70 |
| $k\times10^2$, s$^{-1}$ | 3.22 | 3.26 | 3.21 | 3.23 | 3.29 | 3.18 | 3.19 | 3.22 | 3.16 | 3.24 |
| $T = 308$ K | | | | | | | | | | |
| $A$, μV | 13.58 | 26.35 | 39.06 | 50.30 | 63.37 | 75.39 | 87.48 | 102.38 | 113.03 | 122.14 |
| $k\times10^2$, s$^{-1}$ | 3.42 | 3.40 | 3.41 | 3.48 | 3.39 | 3.43 | 3.41 | 3.37 | 3.45 | 3.43 |
| $T = 313$ K | | | | | | | | | | |
| $A$, μV | 13.20 | 24.74 | 37.74 | 46.56 | 58.46 | 70.70 | 81.78 | 91.85 | 103.69 | 112.69 |
| $k\times10^2$, s$^{-1}$ | 3.51 | 3.50 | 3.56 | 3.48 | 3.54 | 3.49 | 3.47 | 3.55 | 3.51 | 3.52 |

As follows from the data of Table 4, the $k$ value is virtually independent on the number density $n_X$ ($X$ = IFs or SWCNTs) at a given temperature, for both IFs and SWCNTs. Table 5 lists the average values of this parameter in function of temperature, with higher $k$ values obtained at higher temperatures. The $A$ value increases linearly with the number density for both IFs and SWCNTs, with the respective linear regressions shown in Figure 8. Note that the intercepts are non-zero, which we explain by the presence of residual ionic species in the water. The slopes $F$ of the respective linear regressions are also listed in Table 5.

Table 5. Average values of the rate parameter $k$, and of the slope $F$ of the regression plots of the signal amplitude $A$ on the number density $n_X$ of the studied species ($X$ = IFs or SWCNTs), in function of temperature.

| | IFs | | SWCNTs | |
|---|---|---|---|---|
| $T$, K | $k$, s$^{-1}$ | $F$, μV cm$^3$ | $k$, s$^{-1}$ | $F$, μV cm$^3$ |
| 278 | 0.0089±0.0001 | 6.19±0.12 | 0.0281±0.0004 | 40.72±0.56 |
| 288 | 0.0097±0.0002 | 5.65±0.11 | 0.0302±0.0005 | 37.94±0.49 |
| 298 | 0.0105±0.0002 | 5.20±0.12 | 0.0324±0.0004 | 35.58±0.73 |
| 308 | 0.0115±0.0003 | 4.80±0.11 | 0.0342±0.0007 | 33.51±0.61 |
| 313 | 0.0119±0.0003 | 4.63±0.10 | 0.0351±0.0006 | 32.61±0.47 |

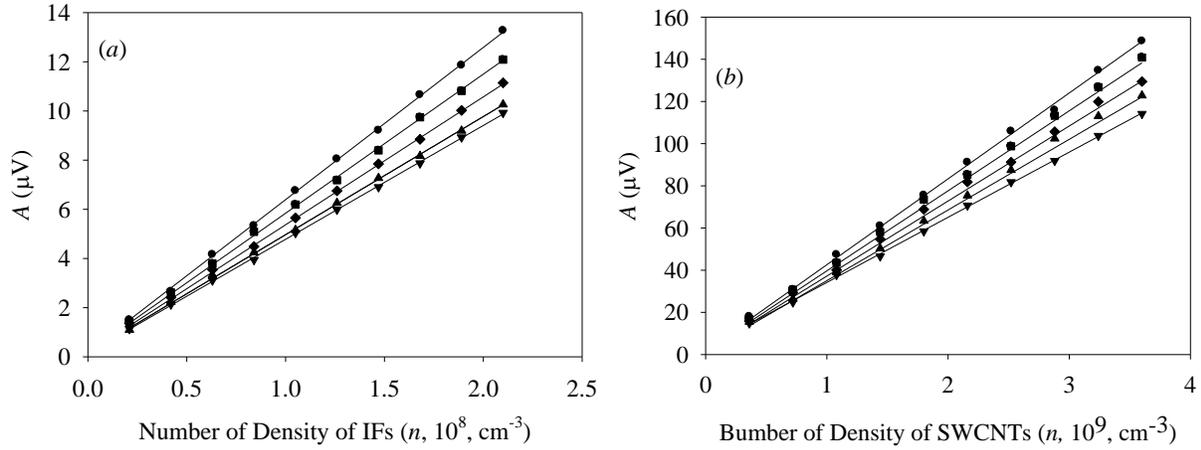

Figure 8. Dependences of $A$ on $n_X$ ($X$ = IFs and SWCNTs) for different temperatures: circle – $T$ = 278 K; square – $T$ = 288 K; diamond – $T$ = 298 K; triangle up – $T$ = 308 K; triangle down – $T$ = 313 K.

As follows from Figure 8 and Table 5, the regression slope $F$ decreases with increasing temperature. To understand the temperature dependences and calculate the average resistance of a single IF or SWCNT, we need an adequate model describing their interaction with an electrode matrix.

Two processes should be taken into account: (1) diffusion of the $X$ species ($X$ = IF or SWCNT) to the electrode matrix and (2) reversible adsorption of the $X$ species by the electrode matrix surface. In the present study, we analyzed two limiting cases: (*a*) the kinetic limit, when the diffusion rate of $X$ to the matrix surface is much faster than the adsorption and desorption rates, and (*b*) the diffusion limit, when the diffusion is much slower than the adsorption and desorption.

We shall first look at the physics of the resistance measurements. We shall use the following parameters: the resistivity $\rho_X$ [$\Omega \cdot \text{cm}^2 \cdot \text{m}$], distance between the electrodes $h$, average length of $X$ $l_X$, its diameter $d_X$, and $\varphi$, the angle between the axes of the electrode and of the species $X$. The resistance created by a single $X$ absorbed between the two electrodes, averaged over the angle, will be given by:

$$r_X = \rho_X \frac{h}{d_X^2} \int_{\arcsin\left(\frac{h}{l_X}\right)}^{\frac{\pi}{2}} \frac{d\varphi}{\sin(\varphi)} \tag{4}$$

Taking into account the known parameters of the IFs and SWCNTs, the relationship (4) may be rewritten as follows:

$$r_{IF} = 3.76 \times 10^8 \rho_{IF} \, [\Omega]$$
$$r_{SWCNT} = 1.70 \times 10^9 \rho_{SWCNT} \, [\Omega] \tag{5}$$

where the average reduced resistance of both species are the unknown parameters. The total resistivity of the electrode matrix, and therefore the observed signal, is dependent on the surface number density of the adsorbed species. To estimate the absorbed surface number density of the respective species, we carried out analysis of two limits mentioned above.

(*a*) The kinetic limit

The diffusion rate to the electrode matrix surface is much faster than the adsorption and desorption rates at the $z = 0$ coordinate, corresponding to the matrix surface. We introduce the adsorption rate constant $k_1$, and desorption rate constant $k_{-1}$. The sample number density at $t = 0$ is $n_0$. Thus, the equations for the surface number density are:

$$\frac{dn_{surf}}{dt} = -k_{-1} n_{surf} + k_1 n_{liq}$$
$$n_{liq} = n_0 - n_{surf} \tag{6}$$

Their solution is given by

$$n_{surf} = \frac{k_1 n_0}{k_{-1} + k_1} \left(1 - e^{-(k_{-1} + k_1)t}\right) \tag{7}$$

The measured signal may be presented as follows:

$$U_X = V_0 \left( \frac{R_2(R_3 + R_X)}{R_2 R_3 + (R_2 + R_3)R_X} - \frac{R_1}{R_1 + R_4} \right) = \frac{V_0 R}{2(R + 2R_X)} \qquad (8)$$

$$R_X = \frac{r_X}{n_{X,surf}}$$

where $r_X$ is the resistance of a single species attached to the EM and immersed in bidistilled / deionized water. Taking into account that $R_1 = R_2 = R_3 = R_4 = R_0$, we obtain

$$U_X = \frac{V_0 R n_{X,Surf}}{2(R n_{X,Surf} + 2r_X)} \approx \frac{V_0 R n_{X,Surf}}{4 r_X} \qquad (9)$$

and the dependence for the measured signal may be rewritten as follows:

$$U_X = \frac{V_0 R}{4 r_X} \frac{k_1 n_{X,0}}{k_{-1} + k_1} \left( 1 - e^{-(k_{-1} + k_1)t} \right) \qquad (10)$$

The temperature dependence of the parameters of Table 5 may be explained, rewriting relationship (10) in the form:

$$U_X = \frac{V_0 R}{4 r_X} \frac{K_{eq}}{1 + K_{eq}} n_{X,0} \left( 1 - e^{-(k_{-1} + k_1)t} \right) \qquad (11)$$

Thus, we identify $\frac{V_0 R}{4 r_X} \frac{K_{eq}}{1 + K_{eq}} \approx \frac{V_0 R}{4 r_X} K_{eq}$ as the experimental parameter $A$, and $k_{-1} + k_1$ as the experimental parameter $k$. The temperature-dependent contributions are:

$$K_{eq} = e^{-\frac{\Delta G_X}{k_B T}} = e^{-\frac{\Delta H_X - T \Delta S_X}{k_B T}}$$

$$k'_1 = \gamma \frac{D_i(T)}{h} S \qquad (12)$$

$$k_{-1} = C e^{-\frac{E_{a,X}}{k_B T}}$$

where $\Delta G_X$, $\Delta H_X$ and $\Delta S_X$ are the Gibbs free energy, enthalpy and entropy changes in the desorption process, $k_B$ is the Boltzmann constant, $\gamma_I$ is the absorption efficiency of the *i-th* species by the EM, $D_i(T)$ its diffusion coefficient in water, $h$ is characteristic size in the direction perpendicular to the EM surface, $S$ is the EM area free from electrode surfaces and it is about 20% from total EM surface,, $C$ and $E_{a,X}$ are the pre-exponential factor and the activation energy for the activated desorption. We shall assume that $\gamma_i = 1$ for both species. We shall also neglect temperature dependence of $k_1$, due to low activation energy of the diffusion. Using the viscosity coefficient of water [24] and representing the studied species as spherical particles with the effective radius of $z_{eff} = \frac{3}{4\pi}\sqrt[3]{V_i}$, we estimated $k'_1 = 2.90\times10^{-11}$ cm$^3$s$^{-1}$ and $k'_1 = 1.75\times10^{-12}$ cm$^3$s$^{-1}$ for IFs and SWCNTs, respectively. Thus, the plots of $-\ln(k_{-1}) = -\ln(k - k'_1 n_X)$ vs. $1/T$ and of $\ln(F/F_{max})$ vs. $1/T$ are shown in Fig. 9.

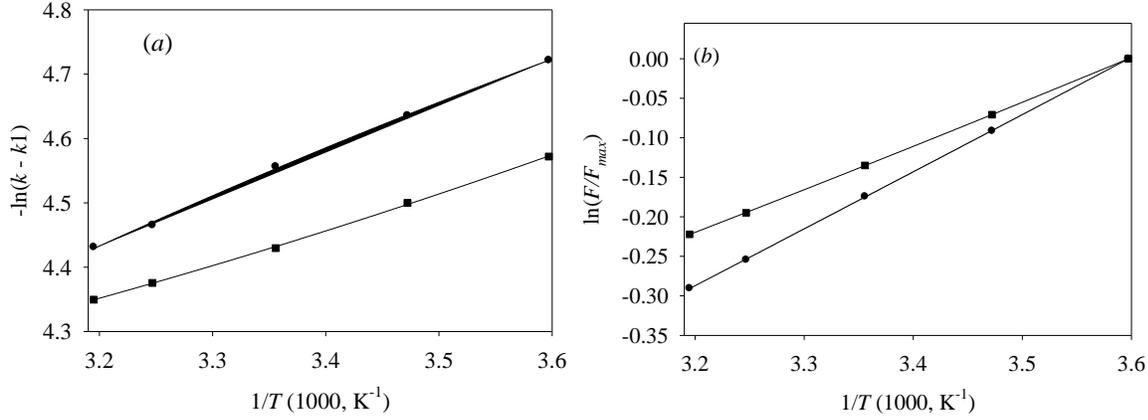

Figure 9. (*a*) Dependences of $-\ln(k)$ vs $1/T$ (a) IFs – $C = 0.122$ s$^{-1}$, $E_a = 482$ cm$^{-1}$; (*b*) SWCNTs – $C = 0.219$ s$^{-1}$, $E_a = 370$ cm$^{-1}$; (*b*) Dependences of $\ln(A/A_{max})$ vs $1/T$: (*a*) IFs - $\Delta H_{IF} = 479$ cm$^{-1}$; SWCNTs - $\Delta H_{SWCNT} = 374$ cm$^{-1}$.

We conclude using the data of Fig. 9 and the parameter values in its caption that $E_{a,X} \approx \Delta H_X$. This is equivalent of temperature-independent $k_1$. We estimated the entropy factor $e^{-\frac{\Delta S}{k_B}}$ for both species taking into account the reduction in the number of available states for the adsorbed $X$ in comparison to dissolved $X$. These estimates produced about $1.7\times10^{-3}$ for both species.

Using the above estimates and the relationship (11), we calculated the $r_X$ value, which is about $(6.5\pm0.4)10^{11}$ Ω and $(3.6\pm0.2)10^{13}$ Ω, respectively for IFs and SWCNTs. Taking into account that average diameters of 10 and 4.7 nm and lengths of 150 and 10 μm, for the IFs and SWCNTs respectively, we calculated the average reduced resistivity of the IFs and SWCNTs of $(3.1\pm0.3)\times10^{-4}$ and $(2.8\pm0.2)\times10^{-4}$ Ω·m$^{-1}$·cm$^2$, respectively. For reference, these values are quite close to the tabulated value for aluminum: $2.8\times10^{-4}$ Ω·m$^{-1}$·cm$^2$ [18]. Thus, the analysis of the experimental data in the kinetic limit produced reduced resistance values comparable to those of metals, for both IF and SWCNT samples. Such results were to be expected for SWCNTs, while coming as some surprise for the IFs. Note, however, that high electric conductivity has been reported earlier for macroscopic samples of some proteins [25]. Therefore, we conclude that the presently studied IFs have the same high values of the electric conductivity as carbon nanotubes.

(*b*) The diffusion limit

Here the-time dependent number density of adsorbed *X* may be obtained analysing the flux towards the electrode matrix surface, taking into account the boundary conditions at the surface. The flux to the surface may be presented as follows:

$$J_{liq} = -D(T)\frac{\partial n_{liq}}{\partial z} \tag{13}$$

where $D(T)$ is the temperature-dependent diffusion coefficient of *X* in water. Boundary condition can be presented as follows:

$$n_{surf}(z=0, t=0) = 0$$
$$n_{surf}(z, t \to \infty) = K_{eq}(T) n_{liq}(z, t \to \infty) \tag{14}$$

In ours analysis, we considered quasistationary conditions ($t \to \infty$), i.e. we analyzed only the parameter *A*. Taking into account that $n_0 = n_{surf} + n_{liq}$, the surface number density of the adsorbed species in given by:

$$n_{surf}(z=0) = \frac{K_{eq}(T) n_0}{1 + K_{eq}(T)} \tag{15}$$

i.e. the relationship (11) may once again be obtained to describe the observed signal. Thus, the same values of the reduced resistance of the IFs and SWCNT as obtained above were calculated here, using the already estimated value of $K_{eq}$.

c) Axial light transmission

We measured light transmission by the IFs, using the capillary matrix with 15 nm holes, and by the SWCNTs, using the matrix with 5 nm holes. The experimental procedure was as follows. First, the required capillary matrix was mounted into the Pyrex cell (Figure 3). Both halves of the cell were filled with 3% aqueous NaCl solution. Hg lamp light passing through appropriate filters and entering one of the cell windows and illuminated the capillary matrix, with the system background recorded for 5 mins. The lamp shutter was closed, and the left section of the cell cleaned and filled with the sample suspension, diluted in water at the ratio from $10^3$ to $10^4$. Finally, the lamp shutter was opened and the signal dynamics recorded.

A flux of water through the capillary matrix was created by the osmosis in these experiments, with the sample species pulled along by the flux and into the holes. The holes became filled and clogged with the sample species, which should change the optical transmission of the capillary matrix, provided the earlier developed models are correct [4-6]. All of the light transmission data are presented with the background value subtracted. Note that typical background count rate was ca. 170 $s^{-1}$. Figure 10 shows typical examples of the transmission signal dynamics for both samples. At each of the probing radiation wavelengths, the signal was normalized to the probing light intensity and corrected for the wavelength-dependent sensitivity curve of the PMT.

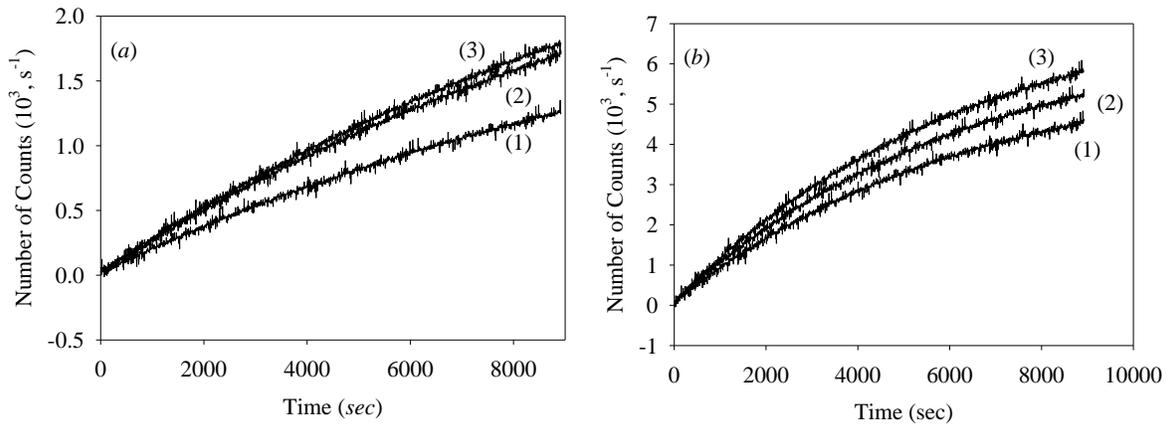

Figure 10. Dependence of the number of PMT counts vs. time measured for radiation light wavelengths of (1) 404.7 nm, (2) 435.8 nm and (3) 546.1 nm, respectively; at number of density of Ifs of (*a*) $2.1 \times 10^5$ $cm^{-3}$ IFs, and (*b*) $3.6 \times 10^5$ $cm^{-3}$ SWCNTs. System temperature is 288 K.

The kinetics of Fig. 10 were fitted by an exponential function, with the values of the fitting parameters of $B$ and $k_{em}$ presented in Table 6:

$$I_{em}(t) = B\left(1 - e^{-k_{em}t}\right) \tag{16}$$

Table 6. The values of the fitting parameters $B$ (photon counting rate) and $k_{em}$.

|  | IFs | | SWCNTs | |
|---|---|---|---|---|
| Wavelength, nm | $B$, $s^{-1}$ | $k_{em} \times 10^5$, $s^{-1}$ | $B$, $s^{-1}$ | $k_{em} \times 10^4$, $s^{-1}$ |
| 404.7 | 3612 | 7.58 | 7631 | 1.58 |
| 435.8 | 3431 | 7.65 | 6830 | 1.62 |
| 546.1 | 2523 | 7.38 | 5923 | 1.51 |

Since the signal was normalized to the lamp intensity and the PMT sensitivity, the relative quantum yields of the light energy transmission at different wavelengths by the IFs and SWCNTs may be determined as the ratio of the respective values of *B*.

These results may be explained taking into account the filling dynamics of the matrix capillary by IFs or SWCNTs, their excitation in the input (left) section of the cell, and re-emission electronic excitation energy in the output (right) section of the cell. A detailed theoretical analysis of the excitation energy transfer by IFs and CNTs was presented earlier [4,5], with the present experimental results demonstrating the validity of the theory.

Time of electronic excitation transmission along IFs and SWCNTs was measured using laser pulses at 480 nm (fundamental of the dye laser using Coumarine-102) to excite IFs and 700 nm (fundamental of the dye laser using Nile blue) to excite SWCNTs. The recorded waveforms are shown in Fig. 11.

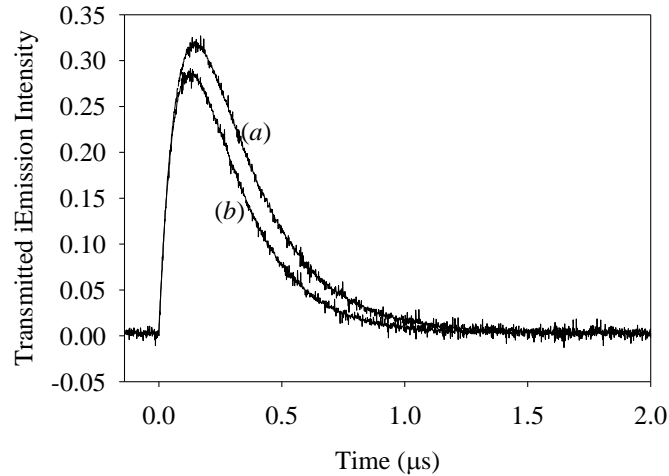

Figure 11. Excitation energy transfer dfynamics along IFs (*a*) and SWCNTs (*b*).

These waveforms were fitted by the function

$$I(t) = A\left(e^{-\frac{t}{\tau_1}} - e^{-\frac{t}{\tau_2}}\right) \qquad (17)$$

where $A$ is the signal amplitude, and $\tau_i$ the characteristic times of signal buildup and decay, with the parameter values listed in Table 7.

<Insert Table 7>

Table 7. Characteristic times of the transmitted emission buildup and decay for the IF and SWCNT samples.

| Sample | $\tau_1$, ns | $\tau_2$, ns |
| --- | --- | --- |
| IF | 53.1±2.1 | 231.9±7.1 |
| SWCNT | 49.3±1.3 | 183.4±6.8 |

The buildup time $\tau_1$ is quite similar for the two species, probably due to the similar mechanism of the excitation transport along the respective nanostructures. This mechanism apparently involves in both species the participation of a delocalized π-conjugated system, defining their properties. However, taking into account the significant difference in the length of the IFs and SWCNTs, we would expect significant differences in the buildup time that should depend of the excitation transport velocity along these species and their length, and thus probably longer for the IFs. The $\tau_2$ decay time is slightly longer for the IFs as compared to SWCNTs, however, for both samples these times are significantly shorter than those recorded in bulk samples (see Table 3). This result may be explained by the excited state relaxation caused by the interaction with the Si capillary wall, with the semiconducting Si efficiently interacting with the delocalized wavefunction of the excited state, or by the stimulated emission (superemission) generated along the axis of the IFs or SWCNTs [23].

The observed optical effects for SWCNTs are not unexpected, in view of the data reported by different authors earlier [23,26-29]. However, optical properties of IFs of Müller cells, or any other cells, have never been reported yet. Many studies analyze the IF structure and their cytoskeletal function [30-34], while their other biological functions were only explored quite recently [35-37]. The current study is the first one reporting direct measurements of porcine IFs optical properties and electric conductivity. The presently obtained results thus demonstrate the correctness of the fundamental assumptions made earlier [6], showing that IFs should be important for the light energy transmission in an inverted retina, and confirming the earlier developed theoretical approach [4-6].

## IV. Conclusions

Thus, the present report provides an experimental confirmation of the earlier postulated properties [3-6] of electric conductivity, light energy transmission and spectral selectivity of the IFs and CNTs. We report experimental and theoretical methods for measuring and calculating the reduced resistivity of the IFs ($\rho_{IF} = (3.1\pm0.3)\times10^{-4}$ $\Omega \cdot m^{-1} \cdot cm^{2}$) and SWCNTs ($\rho_{SWCNT} = (2.8\pm0.2)\times10^{-4}$ $\Omega \cdot m^{-1} \cdot cm^{2}$). We also report the experimental method to study the light energy transmission of the IFs and SWCNTs. We found that both IFs and SWCNTs transmit light energy along their axis, in a good agreement with the earlier developed theoretical models [4-6]. Therefore, the earlier proposed mechanism [3-6] of high-contrast vision in an inverted retina was confirmed by the presently reported experimental data.

**Acknowledgements**: V.M. is grateful to NASA, grant PR NASA EPSCoR NNX13AB22A, for financial support.